\documentclass[reqno,12pt,a4paper]{amsart}

\usepackage{mathrsfs}
\usepackage{amssymb}
\usepackage{amsfonts}
\usepackage{latexsym}
\usepackage{amsthm}
\usepackage[dvips]{color}

\usepackage{graphicx}
\def\lb{\label}

\newcommand{\er}[1]{\textrm{(\ref{#1})}}

\begin{document}


\renewcommand{\theequation}{\arabic{section}.\arabic{equation}}
\theoremstyle{plain}
\newtheorem{theorem}{\bf Theorem}[section]
\newtheorem{lemma}[theorem]{\bf Lemma}
\newtheorem{corollary}[theorem]{\bf Corollary}
\newtheorem{proposition}[theorem]{\bf Proposition}
\newtheorem{definition}[theorem]{\bf Definition}
\newtheorem{remark}[theorem]{\it Remark}

\def\a{\alpha}  \def\cA{{\mathcal A}}     \def\bA{{\bf A}}  \def\mA{{\mathscr A}}
\def\b{\beta}   \def\cB{{\mathcal B}}     \def\bB{{\bf B}}  \def\mB{{\mathscr B}}
\def\g{\gamma}  \def\cC{{\mathcal C}}     \def\bC{{\bf C}}  \def\mC{{\mathscr C}}
\def\G{\Gamma}  \def\cD{{\mathcal D}}     \def\bD{{\bf D}}  \def\mD{{\mathscr D}}
\def\d{\delta}  \def\cE{{\mathcal E}}     \def\bE{{\bf E}}  \def\mE{{\mathscr E}}
\def\D{\Delta}  \def\cF{{\mathcal F}}     \def\bF{{\bf F}}  \def\mF{{\mathscr F}}
\def\c{\chi}    \def\cG{{\mathcal G}}     \def\bG{{\bf G}}  \def\mG{{\mathscr G}}
\def\z{\zeta}   \def\cH{{\mathcal H}}     \def\bH{{\bf H}}  \def\mH{{\mathscr H}}
\def\e{\eta}    \def\cI{{\mathcal I}}     \def\bI{{\bf I}}  \def\mI{{\mathscr I}}
\def\p{\psi}    \def\cJ{{\mathcal J}}     \def\bJ{{\bf J}}  \def\mJ{{\mathscr J}}
\def\vT{\Theta} \def\cK{{\mathcal K}}     \def\bK{{\bf K}}  \def\mK{{\mathscr K}}
\def\k{\kappa}  \def\cL{{\mathcal L}}     \def\bL{{\bf L}}  \def\mL{{\mathscr L}}
\def\l{\lambda} \def\cM{{\mathcal M}}     \def\bM{{\bf M}}  \def\mM{{\mathscr M}}
\def\L{\Lambda} \def\cN{{\mathcal N}}     \def\bN{{\bf N}}  \def\mN{{\mathscr N}}
\def\m{\mu}     \def\cO{{\mathcal O}}     \def\bO{{\bf O}}  \def\mO{{\mathscr O}}
\def\n{\nu}     \def\cP{{\mathcal P}}     \def\bP{{\bf P}}  \def\mP{{\mathscr P}}
\def\r{\rho}    \def\cQ{{\mathcal Q}}     \def\bQ{{\bf Q}}  \def\mQ{{\mathscr Q}}
\def\s{\sigma}  \def\cR{{\mathcal R}}     \def\bR{{\bf R}}  \def\mR{{\mathscr R}}
\def\S{\Sigma}  \def\cS{{\mathcal S}}     \def\bS{{\bf S}}  \def\mS{{\mathscr S}}
\def\t{\tau}    \def\cT{{\mathcal T}}     \def\bT{{\bf T}}  \def\mT{{\mathscr T}}
\def\f{\phi}    \def\cU{{\mathcal U}}     \def\bU{{\bf U}}  \def\mU{{\mathscr U}}
\def\F{\Phi}    \def\cV{{\mathcal V}}     \def\bV{{\bf V}}  \def\mV{{\mathscr V}}
\def\P{\Psi}    \def\cW{{\mathcal W}}     \def\bW{{\bf W}}  \def\mW{{\mathscr W}}
\def\o{\omega}  \def\cX{{\mathcal X}}     \def\bX{{\bf X}}  \def\mX{{\mathscr X}}
\def\x{\xi}     \def\cY{{\mathcal Y}}     \def\bY{{\bf Y}}  \def\mY{{\mathscr Y}}
\def\X{\Xi}     \def\cZ{{\mathcal Z}}     \def\bZ{{\bf Z}}  \def\mZ{{\mathscr Z}}
\def\O{\Omega}

\newcommand{\gA}{\mathfrak{A}}
\newcommand{\gB}{\mathfrak{B}}
\newcommand{\gC}{\mathfrak{C}}
\newcommand{\gD}{\mathfrak{D}}
\newcommand{\gE}{\mathfrak{E}}
\newcommand{\gF}{\mathfrak{F}}
\newcommand{\gG}{\mathfrak{G}}
\newcommand{\gH}{\mathfrak{H}}
\newcommand{\gI}{\mathfrak{I}}
\newcommand{\gJ}{\mathfrak{J}}
\newcommand{\gK}{\mathfrak{K}}
\newcommand{\gL}{\mathfrak{L}}
\newcommand{\gM}{\mathfrak{M}}
\newcommand{\gN}{\mathfrak{N}}
\newcommand{\gO}{\mathfrak{O}}
\newcommand{\gP}{\mathfrak{P}}
\newcommand{\gQ}{\mathfrak{Q}}
\newcommand{\gR}{\mathfrak{R}}
\newcommand{\gS}{\mathfrak{S}}
\newcommand{\gT}{\mathfrak{T}}
\newcommand{\gU}{\mathfrak{U}}
\newcommand{\gV}{\mathfrak{V}}
\newcommand{\gW}{\mathfrak{W}}
\newcommand{\gX}{\mathfrak{X}}
\newcommand{\gY}{\mathfrak{Y}}
\newcommand{\gZ}{\mathfrak{Z}}

\def\ve{\varepsilon}   \def\vt{\vartheta}    \def\vp{\varphi}    \def\vk{\varkappa}

\def\Z{{\mathbb Z}}    \def\R{{\mathbb R}}   \def\C{{\mathbb C}}    \def\K{{\mathbb K}}
\def\T{{\mathbb T}}    \def\N{{\mathbb N}}   \def\dD{{\mathbb D}}


\def\la{\leftarrow}              \def\ra{\rightarrow}            \def\Ra{\Rightarrow}
\def\ua{\uparrow}                \def\da{\downarrow}
\def\lra{\leftrightarrow}        \def\Lra{\Leftrightarrow}


\def\lt{\biggl}                  \def\rt{\biggr}
\def\ol{\overline}               \def\wt{\widetilde}
\def\no{\noindent}


\let\ge\geqslant                 \let\le\leqslant
\def\lan{\langle}                \def\ran{\rangle}
\def\/{\over}                    \def\iy{\infty}
\def\sm{\setminus}               \def\es{\emptyset}
\def\ss{\subset}                 \def\ts{\times}
\def\pa{\partial}                \def\os{\oplus}
\def\om{\ominus}                 \def\ev{\equiv}
\def\iint{\int\!\!\!\int}        \def\iintt{\mathop{\int\!\!\int\!\!\dots\!\!\int}\limits}
\def\el2{\ell^{\,2}}             \def\1{1\!\!1}
\def\sh{\sharp}
\def\wh{\widehat}
\def\bs{\backslash}

\def\all{\mathop{\mathrm{all}}\nolimits}
\def\Area{\mathop{\mathrm{Area}}\nolimits}
\def\arg{\mathop{\mathrm{arg}}\nolimits}
\def\const{\mathop{\mathrm{const}}\nolimits}
\def\det{\mathop{\mathrm{det}}\nolimits}
\def\diag{\mathop{\mathrm{diag}}\nolimits}
\def\diam{\mathop{\mathrm{diam}}\nolimits}
\def\dim{\mathop{\mathrm{dim}}\nolimits}
\def\dist{\mathop{\mathrm{dist}}\nolimits}
\def\Im{\mathop{\mathrm{Im}}\nolimits}
\def\Iso{\mathop{\mathrm{Iso}}\nolimits}
\def\Ker{\mathop{\mathrm{Ker}}\nolimits}
\def\Lip{\mathop{\mathrm{Lip}}\nolimits}
\def\rank{\mathop{\mathrm{rank}}\limits}
\def\Ran{\mathop{\mathrm{Ran}}\nolimits}
\def\Re{\mathop{\mathrm{Re}}\nolimits}
\def\Res{\mathop{\mathrm{Res}}\nolimits}
\def\res{\mathop{\mathrm{res}}\limits}
\def\sign{\mathop{\mathrm{sign}}\nolimits}
\def\span{\mathop{\mathrm{span}}\nolimits}
\def\supp{\mathop{\mathrm{supp}}\nolimits}
\def\Tr{\mathop{\mathrm{Tr}}\nolimits}
\def\BBox{\hspace{1mm}\vrule height6pt width5.5pt depth0pt \hspace{6pt}}
\def\where{\mathop{\mathrm{where}}\nolimits}
\def\as{\mathop{\mathrm{as}}\nolimits}


\newcommand\nh[2]{\widehat{#1}\vphantom{#1}^{(#2)}}
\def\dia{\diamond}

\def\Oplus{\bigoplus\nolimits}



\def\qqq{\qquad}
\def\qq{\quad}
\let\ge\geqslant
\let\le\leqslant
\let\geq\geqslant
\let\leq\leqslant
\newcommand{\ca}{\begin{cases}}
\newcommand{\ac}{\end{cases}}
\newcommand{\ma}{\begin{pmatrix}}
\newcommand{\am}{\end{pmatrix}}
\renewcommand{\[}{\begin{equation}}
\renewcommand{\]}{\end{equation}}
\def\eq{\begin{equation}}
\def\qe{\end{equation}}
\def\[{\begin{equation}}
\def\bu{\bullet}

\title[{Sharp spectral estimates}]
        {A note on sharp spectral estimates for periodic Jacobi matrices}
\date{\today}

\def\Wr{\mathop{\rm Wr}\nolimits}
\def\BBox{\hspace{1mm}\vrule height6pt width5.5pt depth0pt \hspace{6pt}}

\def\Diag{\mathop{\rm Diag}\nolimits}

\date{\today}

\author
{Anton A. Kutsenko}

\address{Jacobs University (International University Bremen), 28759 Bremen, Germany; email: akucenko@gmail.com}



\begin{abstract}
The spectrum of three-diagonal self-adjoint $p$-periodic Jacobi matrix with positive off-diagonal elements $a_n$ an real diagonal elements $b_n$ consist of intervals separated by $p-1$ gaps $\g_i$, where some of the gaps can be degenerated. The following estimate is true
$$
 \sum_{i=1}^{p-1}|\g_i|\geq\max(\max(4(a_1...a_p)^{\frac1p},2\max a_n)-4\min a_n,\max b_n-\min b_n).
$$
We show that for any $p\in\N$ there are Jacobi matrices of minimal period $p$ for which the spectral estimate is sharp. 
The estimate is sharp for both: strongly  and weakly oscillated $a_n$, $b_n$. Moreover, it improves some recent spectral estimates.
\end{abstract}

\maketitle

\section{Introduction}

The periodic Jacobi matrices corresponds to the finite-difference approximation of second-order partial differential operators with periodic coefficients, e.g. periodic Schr\"odinger operators. The $p$-periodic Jacobi matrix $J\ev J(a,b)$ is the self-adjoint operator acting on $\ell^2(\Z)$ of the form
$$
 (Jy)_n=a_{n-1}y_{n-1}+b_ny_n+a_ny_{n+1},\ \ y=(y_n)_{n\in\Z}\in\ell^2(\Z),
$$ 
where $a=(a_n)_{n\in\Z}$, $b=(b_n)_{n\in\Z}$ are $p$-periodic sequences of real numbers, $a_{n+p}=a_n$, $b_{n+p}=b_n$ for all $n\in\Z$. We assume also that all $a_n>0$ and, for simplicity, $p\ge2$. It is well-known, see, e.g., \cite{vM,Te}, that the spectrum of $J$ is absolutely continuous and consists of $p$ intervals $\s_i$ separated by $p-1$ gaps $\g_i$. Some of the gaps may be degenerated. Let us denote the distance between maximal and minimal spectral points by $r\ev r(a,b)=\l^{\max}-\l^{\min}$. There is an estimate, see \cite{KKr},
\[\lb{rad}
 r\ge4(a_1...a_p)^{\frac1p}.
\]
There is another estimate for the Lebesgue measure of the spectrum, see, e.g., \cite{L,KKr,DS},
\[\lb{mes}
 \sum_{i=1}^{p}|\s_i|\le4(a_1...a_p)^{\frac1p}
\]
Some similar estimates were obtained also for quasi-periodic cases in \cite{PR}. It is not possible to combine \er{rad} and \er{mes} to obtain some non-trivial estimates for the lengths of spectral gaps. For a long time, \er{mes} was
 the best estimate, while in \cite{Ku1} it was improved
\[\lb{mes1}
 \sum_{i=1}^{p}|\s_i|\le 4\min a_n.
\] 
Although \er{mes1} improves significantly \er{mes}, it was very hard to publish it somewhere.
Now, it is possible to combine \er{rad} and \er{mes1} to obtain the estimate for the Lebesgue measure of the spectral gaps
\[\lb{est}
 \sum_{i=1}^{p-1}|\g_i|=r-\sum_{i=1}^{p}|\s_i|\ge4(a_1...a_p)^{\frac1p}-4\min a_n.
\]
The estimate \er{est} is sharp in the following sense.
\begin{theorem}\lb{T1}Let $p\in\N$, $b=(0)_{n\in\Z}$, $a=(1-c\d_{np})_{n\in\Z}$, where $c>0$ and $\d_{np}=1$ if  $n=mp$ for $m\in\Z$ and $\d_{np}=0$ otherwise. Then, for the gaps $\g_i$ of Jacobi matrix $J(a,b)$, we have
\[\lb{estt}
  \sum_{i=1}^{p-1}|\g_i|=\frac{4(p-1)}{p}c+o(c),\ \ \ 4(a_1...a_p)^{\frac1p}-4\min a_n=\frac{4(p-1)}{p}c+o(c).
\] 
Hence, \er{est} is sharp for small $c\to+0$. Moreover, the minimal period of $a$ is $p$.
\end{theorem}

Let $J=J(a,b)$ be some periodic Jacobi matrix. We can always shift the spectrum of $J$ to have $\l^{\max}=-\l^{\min}$. Let $\l\in\R$ be such that $J(a,b+\l(1))=J+\l\cI$ have the property $\l^{\max}=-\l^{\min}$, where $\cI$ is the identity operator. It is easy to see that $\l^{\max}\ge \max_ia_i$, since $\l^{\max}=\sup\frac{\|Jh\|}{\|h\|}$ for $h\in\ell^2(\Z)$ and we can always choose a unit vector $h$  for which $\|Jh\|\ge \max a_n$. Hence $r\ge2\max_ia_i$ and we can write
\[\lb{est2}
 \sum_{i=1}^{p-1}|\g_i|=r-\sum_{i=1}^{p}|\s_i|\ge2\max a_n-4\min a_n,
\]
see \er{mes1}, \er{est}, since the shifting $J+\l\cI$ does not change the off-diagonal elements $a$ and the spectral lengths $|\g_i|$, $|\s_i|$, $r$. Estimate \er{est2} is sharp for some strongly oscillating large $a$ (and $b=0$), where only one $a_i$ is large while the other are small. This is because $\l^{\max}\sim\max_ia_i$ for such $a$ (and $b=0$). But, \er{est2} is useless for slightly oscillating $a$, since RHS in \er{est2} is negative for such $a$.

Let us modify \er{est} as follows:
$$
 \sum_{i=1}^{p-1}|\g_i|\ge4(a_1...a_p)^{\frac1p}-4\min a_n\ge4(\min a_n)^{\frac{p-1}p}\lt((\max a_n)^{\frac{1}p}-(\min a_n)^{\frac{1}p}\rt)\ge
$$
$$
 \frac4p\lt(\frac{\min a_n}{\max a_n}\rt)^{\frac{p-1}p}\lt(\max a_n-\min a_n\rt)\ge\frac2p\lt(\max a_n-\min a_n\rt),
$$
for $p\ge2$, where in the last inequality we use
$$
 \sum_{i=1}^{p-1}|\g_i|\ge\max a_n-\min a_n\ \ {\rm for}\ \ \max a_n\ge4\min a_n,
$$
see \er{est2}. Hence, we obtain
\[\lb{est4}
 \sum_{i=1}^{p-1}|\g_i|\ge\frac2p\lt(\max a_n-\min a_n\rt),
\]
which improves a little bit one of the estimates from \cite{G1}, namely
$$
 p^2\sqrt{p}\max_i|\g_i|\ge\max a_n-\min a_n.
$$ 
Note that two-sided spectral estimates from \cite{G1} supplement essentially the Borg-type results on the existence of spectral gaps, see also \cite{CGR,KKu1}.

Finally, note the estimate, see (1.4) in \cite{KKr}
\[\lb{estb}
 \sum_{i=1}^{p-1}|\g_i|\ge\max b_n-\min b_n
\]
which is obviously sharp for a weakly and strongly oscillated $b$  when $a$ tends to $0$. Combining \er{est}, \er{est2}, and \er{estb} we obtain the estimate announced in the Abstract 
\[\lb{estc}
 \sum_{i=1}^{p-1}|\g_i|\geq\max(\max(4(a_1...a_p)^{\frac1p},2\max a_n)-4\min a_n,\max  b_n-\min  b_n).
\]
Following the discussion above and the results of Theorem \ref{T1}, we conclude that \er{estc} is sharp for weakly and strongly oscillated $a$ and $b$ of arbitrary minimal periods $p\in\N$.

\section{Proof of Theorem \ref{T1}}
Let $J=J(a,b)$ be $p$-periodic Jacobi matrix.
Let us define the matrix
$$
 J(a,b,e^{ik})=\ma b_1 & a_1 & 0 & ... & e^{ik}a_p \\ 
                   a_1 & b_2 & a_2 & ... & 0 \\
                   0 & a_2 & b_3 & ... & 0 \\
                   ... & ... & ... & ... & ... \\
                   e^{-ik}a_p & 0 & 0 & ... & b_p 
 \am.
$$
Denote its eigenvalues as $\l_i(k)$, $i=1,...,p$. Let us recall some well-known facts about the spectral properties of scalar periodic Jacobi matrices, see, e.g., \cite{Te}. It is well known that the spectral components of $J$ consist of $\l_i(k)$, i.e. $\s_i=\cup_{k\in[0,\pi]}\{\l_i(k)\}$. It is possible to take the union up to $k=\pi$ instead of $k=2\pi$, since $\l_i(k)$ is symmetric relatively to the center $k=\pi$ of the Brillouin zone $[0,2\pi]$. The values $\l_i(0)$ and $\l_i(\pi)$ are the edges of $\s_i$. Inside $k\in(0,\pi)$ the functions $\l_i(k)$ are strictly monotonic for the case $a_n>0$, $n\in\N$.

Suppose that $p$ is even, the odd $p$ can be treated similarly. To study the spectrum of the Jacobi matrix $J(a,b)$ with $b=(0)_{n\in\Z}$, $a=(1-c\d_{np})_{n\in\Z}$, it is useful to apply the regular perturbation theory, since $c\to0$. Let us start with the unperturbed Jacobi matrix $J^0=J(a^0,b)$, where $a^0=(1)_{n\in\Z}$. The are no gaps in the spectrum of $J^0$, since in fact it is $1$-peridic Jacobi matrix. Nevertheless, the perturbed Jacobi matrix $J(a,b)$ is $p$-periodic. Thus, we should consider $J(a^0,b)$ as $p$-periodic also. Following the general spectral theory for Jacobi matrices, the edges of spectral components $\s^0_i$ 
are eigenvalues of the following two matrices
$$
 J_{-}^0\ev J(a^0,b,1)=\ma 0 & 1 & 0 & ... & 1 \\
           1 & 0 & 1 & ... & 0 \\
           0 & 1 & 0 & ... & 0 \\
           ... & ... & ... & ... & ... \\
           1 & 0 & 0 & ... & 0
  \am,\ \ J^0_{+}=J(a^0,b,-1)=\ma 0 & 1 & 0 & ... & -1 \\
           1 & 0 & 1 & ... & 0 \\
           0 & 1 & 0 & ... & 0 \\
           ... & ... & ... & ... & ... \\
           -1 & 0 & 0 & ... & 0
  \am.
$$
Let us consider the first matrix. The first matrix $J_{-}^0$ has eigenvalues $\l_n^0=2\cos\frac{2\pi n}{p}$, $n=0,...,p/2$. The eigenvalues $\l_0^0=2$ and $\l_{p/2}^0=-2$ are simple eigenvalues with the orthonormal eigenvectors
$$
 v_0=\frac1{\sqrt{p}}(1)_{j=1}^{p},\ \ v_{p/2}=\frac1{\sqrt{p}}((-1)^j)_{j=1}^p.
$$
Other eigenvalues $\l_n^0=2\cos\frac{2\pi n}{p}$, $n=1,...,p/2-1$ are double eigenvalues with the corresponding orthonormal eigenvectors
$$
 v_{n1}=\frac1{\sqrt{p}}(e^{\frac{2\pi i nj}p})_{j=1}^{p},\ \ v_{n2}^{-}=\frac1{\sqrt{p}}(e^{\frac{-2\pi i nj}p})_{j=1}^{p}.
$$
The edges of spectral components $\s_n$ of $J(a,b)$ are eigenvalues of the following matrices
$$
 J_-\ev J(a,b,1)=J_-^0-cJ_1,\ \ J_+\ev J(a,b,-1)=J_+^0+cJ_1,
$$
where
$$
 J_1=\ma 0 & 0 & 0 & ... & 1 \\
           0 & 0 & 0 & ... & 0 \\
           0 & 0 & 0 & ... & 0 \\
           ... & ... & ... & ... & ... \\
           1 & 0 & 0 & ... & 0
  \am.
$$
The first matrix $J_-$ has eigenvalues $\l_n$ that can be computed by using the regular perturbation theory for small parameter $c$, see, e.g. \cite{K}. Namely,
applying the regular perturbation theory for single and double eigenvalues $\l_n^0$, we obtain
$$
 \ca \l_0=\l_0^0+cv_0^*J_1v_0+o(c)=2+\frac{2c}p+o(c),\\ \l_{p/2}=\l_{p/2}^0+v_{p/2}^*J_1v_{p/2}+o(c)=-2-\frac{2c}p+o(c), \ac
$$
($^*$ denotes the conjugation) and
$$
 \ca\l_{n1}=\l_n^0+c\l_{n1}^0+o(c),\\ \l_{n2}=\l_n^0+c\l_{n2}^0+o(c),\ac
$$
where
$\l_{n1}^0$ and $\l_{n2}^0$ are eigenvalues of the matrix $H_n$ defined by
$$
 H_n=\ma (v_{n1})^*J_1v_{n1} & (v_{n1})^*J_1v_{n2} \\ (v_{n2})^*J_1v_{n1} & (v_{n2})^*J_1v_{n2} \am=\frac1p\ma \cos\frac{2\pi n(p-1)}p & 2e^{\frac{2\pi i n(p+1)}p} \\ 2e^{-\frac{2\pi i n(p+1)}p} & \cos\frac{2\pi n(p-1)}p \am.
$$
Hence, the Lebesgue measure of the gap, which appears around the edge $\l_n^0$ is
$$
 |\g_n|=|\l_{n1}-\l_{n2}|=c|\l_{n1}^0-\l_{n2}^0|+o(c)=\frac{4c}{p}+o(c).
$$
Thus, all the gaps related to the spectral edges corresponding to the eigenvalues of $J_-$ have the same size equivalent to $4c/p$ for small $c>0$. The similar calculations allow us to conclude that the gaps related to the spectral edges  corresponding to the eigenvalues of $J_+$ have also the same size equivalent to $4c/p$ for small $c>0$. This leads to
$$
 \sum_{n=1}^{p-1}|\g_n|=\frac{4c(p-1)}{p}+o(c),
$$
which proves first identity in \er{estt}. Second identity in \er{estt} is trivial, since all $a_n=1$ except one $a_n=1-c$.

\end{document}